\documentclass[fleqn,usenatbib]{mnras}
\usepackage{newtxtext,newtxmath}
\usepackage[T1]{fontenc}
\usepackage{ae,aecompl}

\usepackage{graphicx}   
\usepackage{amsmath}    

\title[Inhomogeneous dust eclipses in young stars]{Inhomogeneous dust eclipses in young stars. The case of CQ Tauri}

\author[A. V. Dodin \& E. A. Suslina]{
A. V. Dodin,$^{1}$\thanks{E-mail: dodin\_nv@mail.ru (AVD)}
E. A. Suslina$^{1}$
\\
$^{1}$ Sternberg Astronomical Institute, M.V. Lomonosov Moscow State University, 13, Universitetskij pr., 119234, Moscow, Russia\\
}

\date{Accepted XXX. Received YYY; in original form ZZZ}

\pubyear{2021}

\begin{document}
\label{firstpage}
\pagerange{\pageref{firstpage}--\pageref{lastpage}}
\maketitle

\begin{abstract}
We find that dust clouds which eclipse young stars obscure the stellar disc inhomogeneously. In the particular case of CQ Tau, we find isolated optically thick structures with sizes $\lesssim0.6R_*$ and derive the typical $A_{V}$ gradient in the plane of the sky, finding it as high as a few magnitudes per stellar radius. The large extinction gradients and complex structure of the obscuring clouds lead not only to a noticeable Rossiter--McLaughlin effect, but also to complex and variable shaping of stellar absorption lines.
\end{abstract}




\section{Introduction}

Photometric variability caused by circumstellar dust has been found in stars of different types: in rare cases in main sequence stars \citep{Schmidt2019}, and more often in evolved stars (AGB, post-AGB, R~CrB). The most diverse manifestations of circumstellar dust can be found in pre-main-sequence stars (TTS, HAeBe) that is related to the active gas--dust discs around them. This group includes several types of variable stars: the UX Ori-type stars, which are distinguished by irregular Algol-like variability \citep{WW98}; `dippers' which show a quasi-periodic variability, probably related to a warped protoplanetary disc \citep{BCAC99}, with about 20 per cent of young stars showing this kind of variability \citep{Cody14}; stars with long and deep dust eclipses like those recently observed in RW~Aur and AA~Tau \citep{SchG15, SchnG15, G18, D19}; and stars with irregular small-amplitude dimmings \citep{PK07, Cody14, Siw18, Gahm18}.

Dust eclipses provide an opportunity to characterize the circumstellar dust by relatively simple methods in the optical region: there are numerous works in which the characteristic sizes of dust grains are inferred from a colour behaviour of the star. In particular, a greyer extinction is commonly interpreted as evidence for a larger average grain size \citep{ST89, ST91, B92, F92, CHW03, HC04, MB09, B15, MA15, PG15, R16, SchG15, SchnG15}. However, in the case of inhomogeneous eclipses the extinction law is also flattened \citep{Natta1984, G19}.

By `an inhomogeneous eclipse', we mean that extinction varies across the stallar disc. It is obvious that homogeneous eclipses do not exist, because each eclipse must have an ingress and egress, during which the stellar disc is obscured unevenly. However, we can neglect inhomogeneities if noticeable changes in extinction across the cloud occur on scales much large than the stellar radius. Some information about the extinction gradient in circumstellar clouds can be obtained from the variability rate during the ingress or egress with some assumptions about the tangential velocity of the clouds. A more promising manifestation of dust inhomogeneities is the distortions that they produce in stellar absorption line profiles. These distortions arise due to partial obscuration of the rotating stellar surface, which results in the weakening of some parts of the profiles, corresponding to the radial velocities of the obscured regions. The Rossiter--McLaughlin effect -- rapid changes of the measured radial velocity of the star in eclipsing binaries during the eclipse -- has the same nature. The presence of the Rossiter--McLaughlin effect in the UX Ori-type stars has been suspected by \citet{GP13} as a possible explanation of discrepancies in radial velocity measurements of RZ Psc.
\\*\indent A large number of high-quality spectra have been obtained during studies of magnetic fields in young stars \citep{Al13, Vil19}. These spectra have revealed distortions in LSD profiles for some stars, but the causes had not been determined, beyond suggestions of an association with circumstellar gas.
Examining these archived spectra, we found that these distortions resemble those expected for the Rossiter--McLaughlin effect. The greatest body of observational data was found for CQ Tauri, which shows well-pronounced distortions, and we will restrict ourselves to this case. In this paper we will show that the observed distortions are indeed related to the clumpy structure of the dust environment, and we will obtain quantitative information about spatial scales of the clumps.

%
%
\section{Observational data}

We use archival\footnote{\url{http://www.cadc-ccda.hia-iha.nrc-cnrc.gc.ca}} spectra of CQ Tau, obtained with ESPaDOnS (CFHT) on 2012 December 22, 25, 28, and 29 in the spectral range 3690--10480\,\AA. On each of the four observational nights, 8 spectra with exposures of 750 sec were obtained with a signal-to-noise ratio of about 80. The telescope archive provides spectra, which have been already processed with {\sc upena}\footnote{\url{https://www.cfht.hawaii.edu/Instruments/Upena}}. Additionally, for ease of analysis, we combined spectral orders, simultaneously reducing the spectral resolution to  $R=30\,000$ by convolution with the Gaussian profile. Then, each spectrum was cleaned of telluric lines with {\sc molecfit}\footnote{\url{http://www.eso.org/sci/software/pipelines/skytools/molecfit}} \citep{Kausch2015}. Spectra obtained over one night were averaged, normalized to the continuum level, and corrected for Earth's radial velocity. We also looked for line profile variability within 2 hours of observations for each of the nights, but no significant changes were found, probably due to an insufficient signal-to-noise ratio.

\section{Properties of CQ Tau}
\subsection{Atmospheric parameters}
\label{sect:prop}
Stellar atmospheric parameters have been estimated by \citet{Al13} as $T_{\rm eff}=6750\pm300$\,K, $\log g =4.0$; \citet{Vil19} give a similar value for $T_{\rm eff}=6800\pm290$\,K. We also try to determine the parameters in the LTE approximation, using {\sc sme}\footnote{\url{http://www.stsci.edu/~valenti/sme.html}} \citep{Pi16} with stellar atmosphere models from the MARCS2012 grid \citep{Gus08, Ber12}, and the line list from the VALD\footnote{\url{http://vald.astro.uu.se/}} database \citep{R15}. We found that, in the expected parameter range, the Balmer lines are sensitive only to $T_{\rm eff}$ and lead to value of $T_{\rm eff}=6850\pm100$\,K, which agrees with previous estimates within the uncertainties. Gravity $\log g$ is poorly constrained: the wings of the Paschen lines lead to $\log g=4.3,$ which seems too high for a young star, while the ionization equilibrium of \ion{Fe}{I/II} requires a very low value of $\log g<3.$ It is possible that the \ion{Fe}{ii} lines are strengthened by non-LTE effects (overionization).
Fortunately for our analysis, other lines are weakly dependent on $\log g,$ allowing us to set $\log g=3.5.$ To better reproduce the absorption lines, we also adjusted microturbulent velocity $v_{\rm mic}=2.3$\,km\,s$^{-1}$ and chemical abundances for 7 elements:
$\varepsilon_{\rm Mg} = -4.20,$
$\varepsilon_{\rm Si} = -4.54,$
$\varepsilon_{\rm Ca} = -5.30,$
$\varepsilon_{\rm Ti} = -7.11,$
$\varepsilon_{\rm Fe} = -4.57,$
$\varepsilon_{\rm Ni} = -5.90,$
$\varepsilon_{\rm Ba} = -9.30$
($\varepsilon$ is the logarithm of relative abundance by number). Some elements (Mg, Ca, Ba) are overabundant by $\sim0.3$ dex with respect to solar values, however, we suspect that these deviations and the enhancement of \ion{Fe}{ii} lines are related and probably caused by non-LTE effects.

\subsection{Radial and rotational velocities}\label{sect:rv}
The radial velocity, as well as the line profiles show variability (see Section~\ref{sect:evid}), the study of which is the subject of this work. As a reference value, we chose $V_{\rm r} = 10$\,km\,s$^{-1},$ measured from the spectrum of 2012 December 22. For the rotational broadening, we will apply $v\sin i = 100$\,km\,s$^{-1}$ \citep{Al13}. More accurate estimates for these quantities can be obtained only after clarifying the nature of their variability.

\subsection{Photometric variability}
Available photometric data show that, from the end of the 19th century until the middle of the 20th century, CQ Tau did not show photometric activity, fluctuating near 9 magnitude in the photographic $B$-band \citep{Grinin2008}. Around 1940, the star entered into a phase of irregular photometric variability with dimmings of up to 3 magnitudes. This photometric behaviour, typical for the UX Ori stars, persists to this day.
We do not have simultaneous photometry for the spectral observations, however, from the AAVSO data on 2012 December 19 and 28 the photoelectric $V$ magnitudes were 11.0 and 10.6. The out-of-eclipse $V$ magnitude is about 9 \citep{Shakh05}; therefore, all spectra were taken in the eclipsed state.

\section{Observational evidence for inhomogeneity}
\label{sect:evid}
\begin{figure*}
\begin{center}
\includegraphics[scale=0.52]{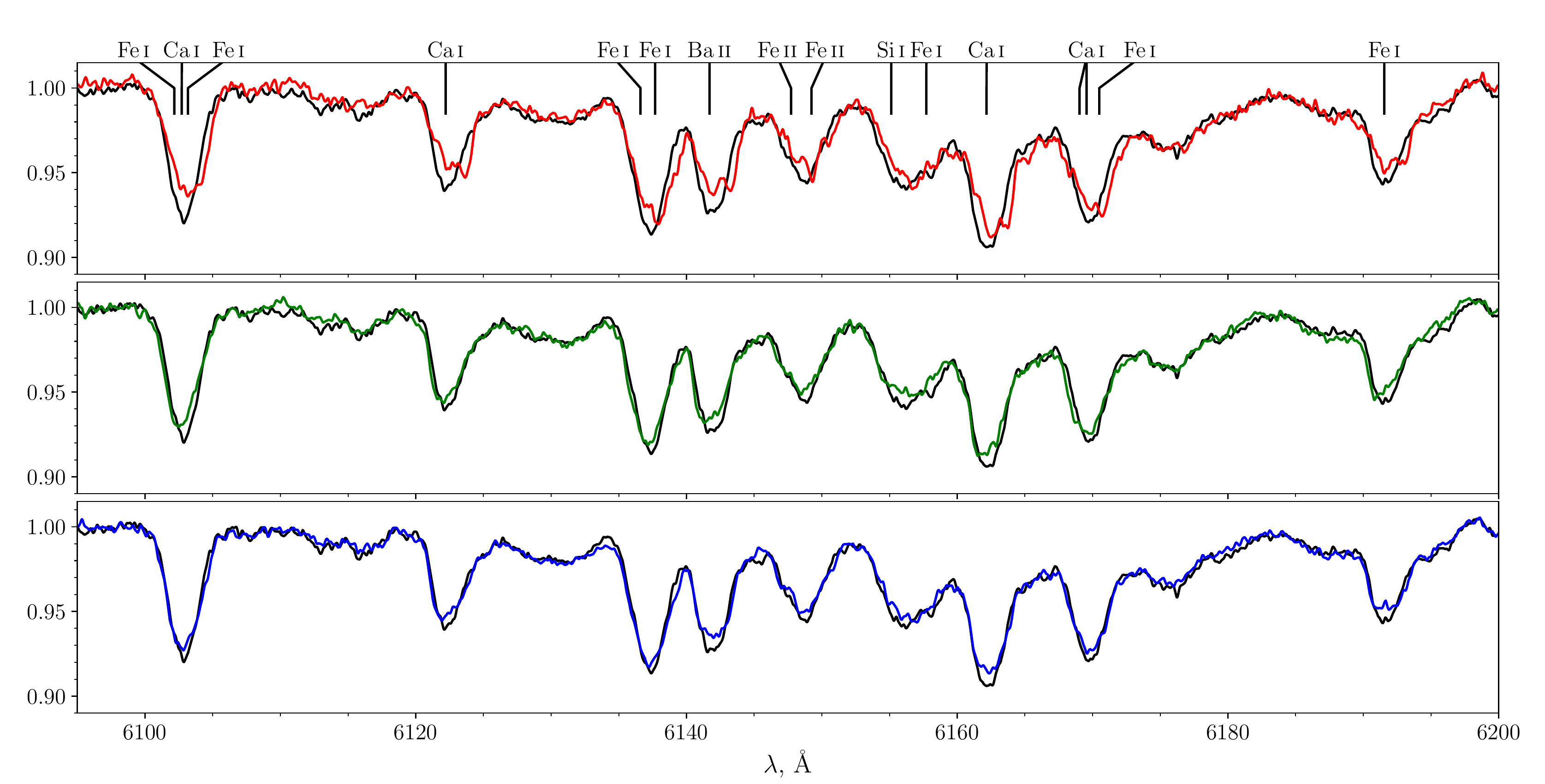}
\caption{Representative parts of CQ Tau spectra. The spectrum obtained on 2012 December 22 is taken as a reference and shown with the black lines. The spectra obtained on 2012 December 25, 28, and 29 are shown with red, green, and blue lines, correspondingly. 
The ions that make the main contribution to the lines are labelled on top.}
\label{fig:spec}
\end{center}
\end{figure*}

Small but representative parts of the CQ Tau spectra are shown in Fig.\,\ref{fig:spec}. It can be seen that all lines have similar distortions, which nearly conserve their equivalent widths (EW), and which do not depend on line properties, although the picture of these distortions is somewhat complicated due to overlapping of the lines. The most impressive distortions were observed on 2012 December 25. Instrumental effects can be ruled out by the absence of these distortions in the telluric lines.

\begin{figure}
\begin{center}
\includegraphics[scale=0.5]{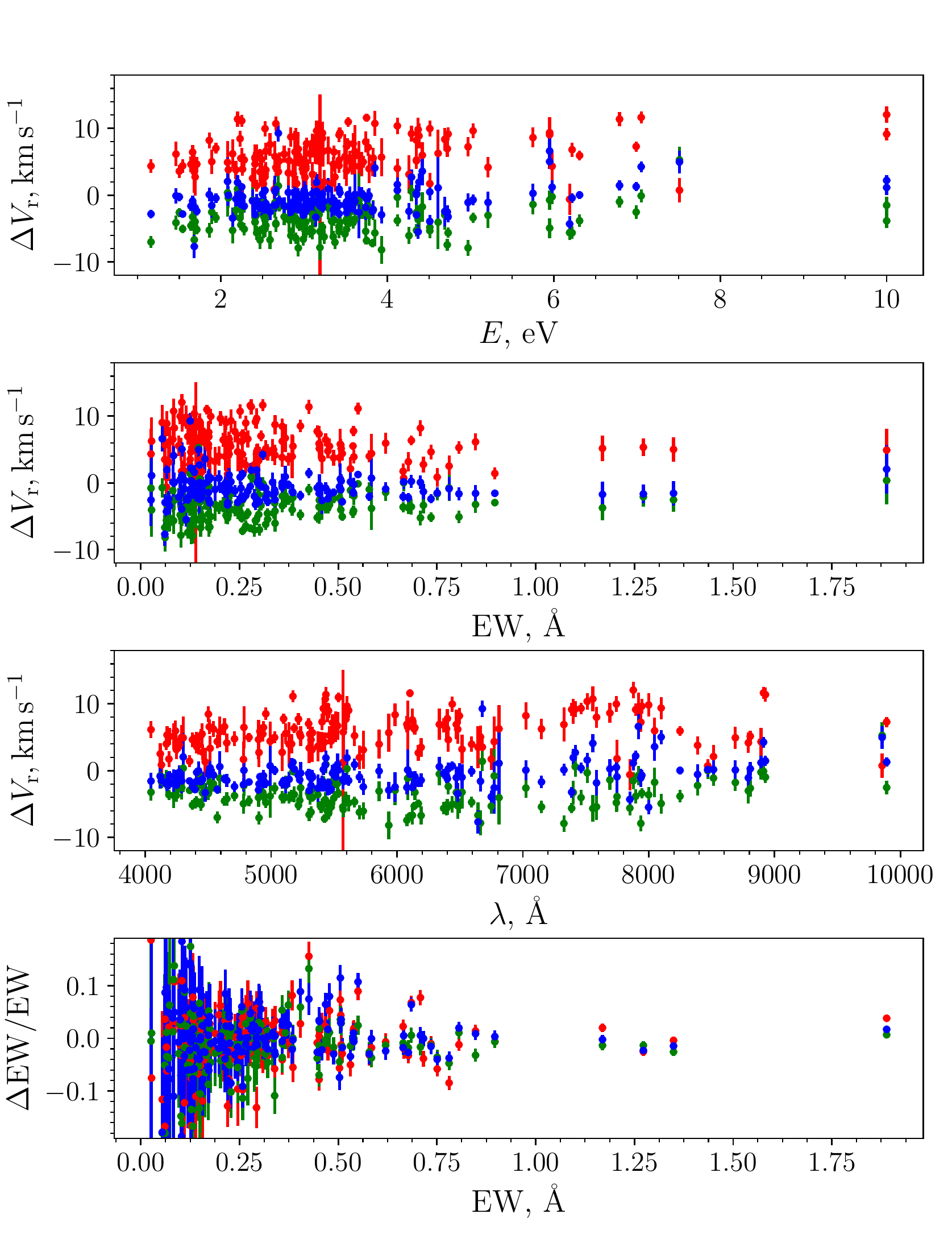}
\caption{ Shift of the radial velocity of individual lines as function of their excitation energy ($E$), equivalent width (EW), and wavelength ($\lambda$). The bottom panel shows relative changes in EW. Colours correspond to Fig.\,\ref{fig:spec}.}
\label{fig:Vrad}
\end{center}
\end{figure}

To describe the observed distortions over the entire spectrum, we selected 146 lines (most of them being blends), for which we determined the radial velocities $\Delta V_{{\rm r}}$ relative to the spectrum taken on 2012 December 22. Fig.\,\ref{fig:Vrad}
shows that the changes in the radial velocities depend weakly on the wavelength, the excitation energy, or the equivalent width, and are reduced mainly to a simple shift, as is expected for an inhomogeneous obscuration of the rotating  stellar surface (see Section\,\ref{sect:model}).
We also show the relative changes in the equivalent widths, which are constant within 10 per cent for ${\rm EW} >0.5$\,\AA. The increasing scatter at low EWs is caused mainly by errors in the continuum level, rather than by physical variability. One can suspect small trends in the presented diagrams, a possible origin of which will be discussed in Section\,\ref{sect:model}.

Because the distortions do not depend on physical parameters of the lines, it is natural to suppose that they are induced by geometrical effects like partial obscuration of the stellar surface.

\section{Simulations}\label{sect:model}
Since the main object of this study is CQ Tau, we perform our simulations for the parameters which best describe the CQ Tau spectrum, however, obtained results can be qualitatively extended to other stars with thin atmospheres.

To simulate distortions of the spectral lines, which appear to be due to partial obscuration, we calculate the specific intensity $I(\lambda, \mu)$ with {\sc sme} v.554 using the parameters of CQ Tau from Section~\ref{sect:prop}. The flux is obtained by integrating the intensities over the stellar disc accounting for the extinction in circumstellar clouds:
\begin{equation}\label{eq:0}
f = \frac{\int I\left[\lambda(1+V/c), \mu\right] e^{-\tau} dS}{\int I_{\rm c}(\lambda, \mu) e^{-\tau} dS}, 
\end{equation}
where $dS$ is a projected area of the surface element; $V=V(x,y)$ is the radial velocity of the element, which for a rigid rotation is $v \sin i x/R_*$;  $\tau = \tau(x,y)$ is the distribution of the optical thickness of the obscuring clouds in the plane of the sky $(x-y)$; $\mu = \mu(x,y)$ is the cosine of the angle between the line of sight and the normal to the stellar surface. The $y$-axis is directed along the projection of the rotation axis; the $x$-axis is orthogonal to $y$.  
It is worth noting that, by denoting $I/I_{\rm c}$ as $u$ and $I_{\rm c}e^{-\tau}/\int{I_{\rm c}e^{-\tau} dS}$ as a weight function $w$, we can rewrite equation\,\ref{eq:0} in the form of a weighted average $f=\int{uwdS},$ i.e. the observed spectrum can be represented as a linear combination of continuum-normalized and Doppler-shifted intensities. 

\begin{figure}
\begin{center}
\includegraphics[scale=0.5]{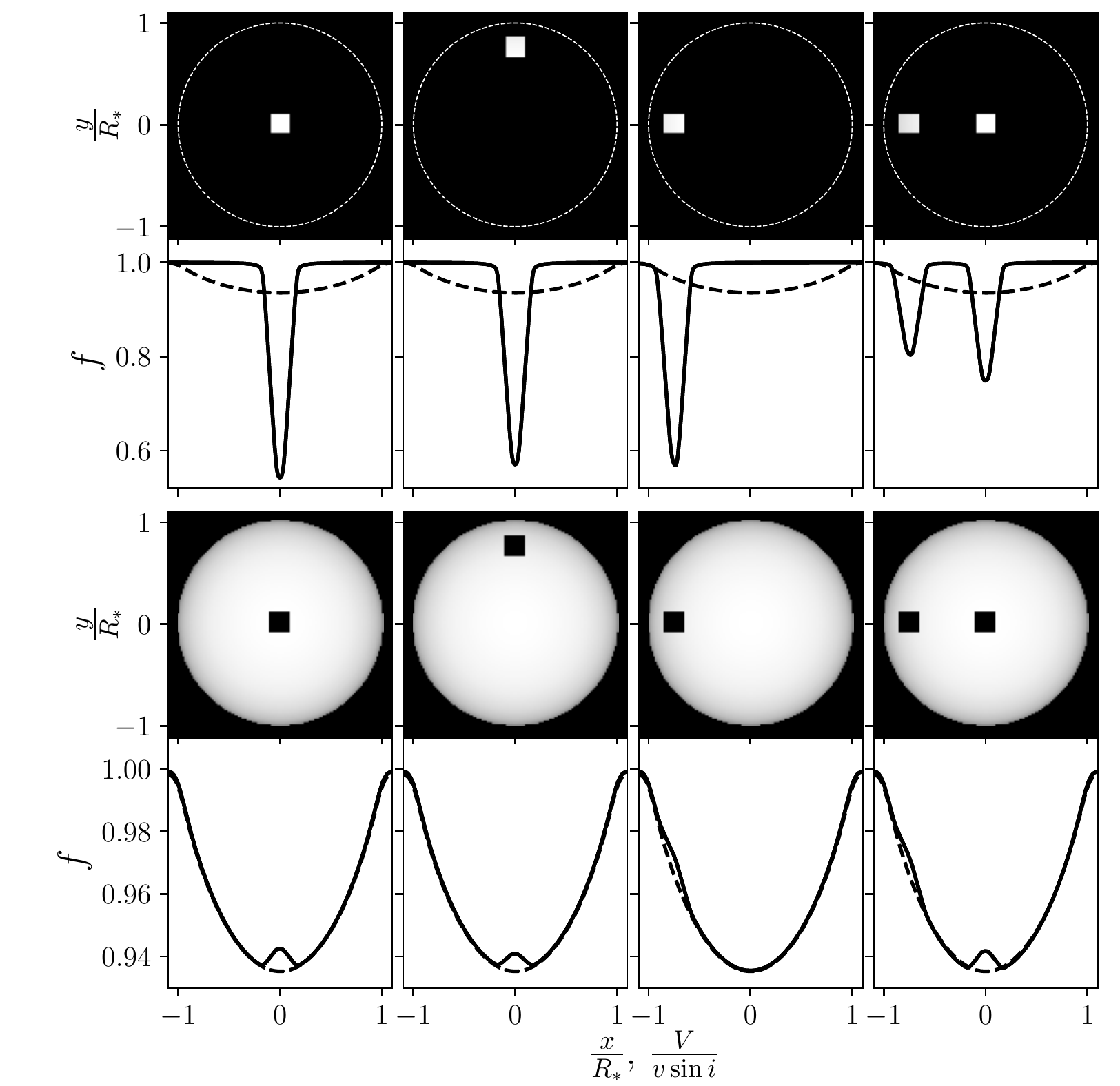}
\caption{Effects of open (the upper panels) or closed (the bottom panels) elements of the stellar surface on the line profile. The dashed line represents the line profile for a clear stellar disc. }
\label{fig:holes}
\end{center}
\end{figure}

To illustrate how the line profile is formed, we show a few simple cases in Fig.\,\ref{fig:holes}, in which small parts of the stellar surface are opened while the rest of the stellar disc is covered by the cloud with infinite $\tau,$ and vice versa. Differences in the limb darkening between the line and continuum lead to the dependence of $I/I_{\rm c}$ on $\mu,$ which makes the equivalent width variable, but maximal changes for the examples in the figure do not exceed 10 per cent in the case of open squares in the upper panels. In more realistic cases, where many angles $\mu$ contribute to the profile, changes of EW will be much less.

\begin{figure}
\begin{center}
\includegraphics[scale=0.6]{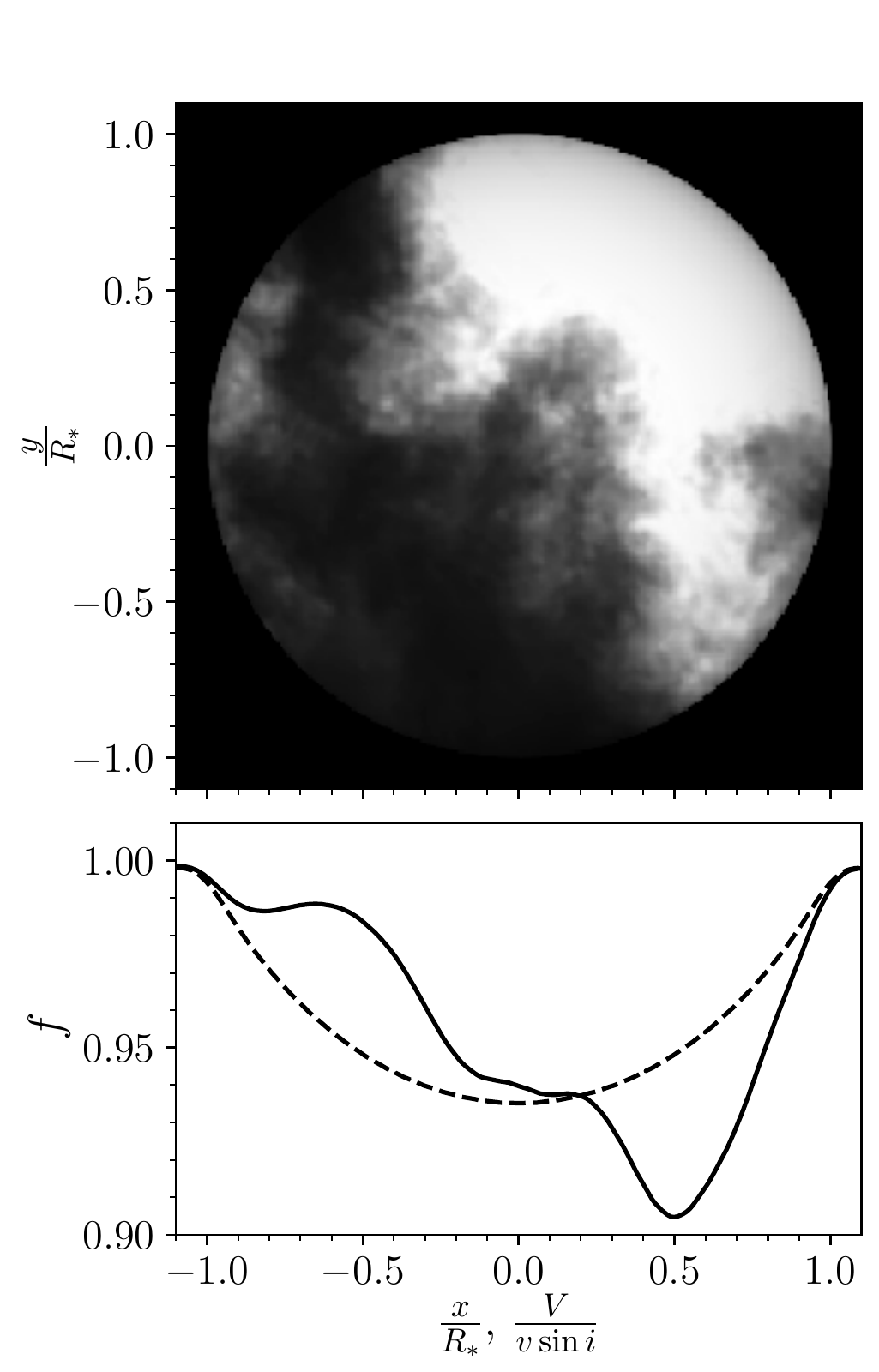}
\caption{A simulated image of the stellar disc (with limb darkening), covered by an inhomogeneous dust cloud and the corresponding line profile. The dashed line represents the out-of-eclipse profile. The maximum optical thickness of the cloud is $\tau_{\rm max}=3.$}
\label{fig:cloud}
\end{center}
\end{figure}

An example of the line profile for the particular case of inhomogeneous obscuration is shown in Fig.\,\ref{fig:cloud}. The profile takes a complex jagged shape, approximately conserving its equivalent width. This is precisely the kind of distortion, that we observe in CQ Tau. The most expressive example is the spectrum taken on 2012 December 25.

\begin{figure}
\begin{center}
\includegraphics[scale=0.6]{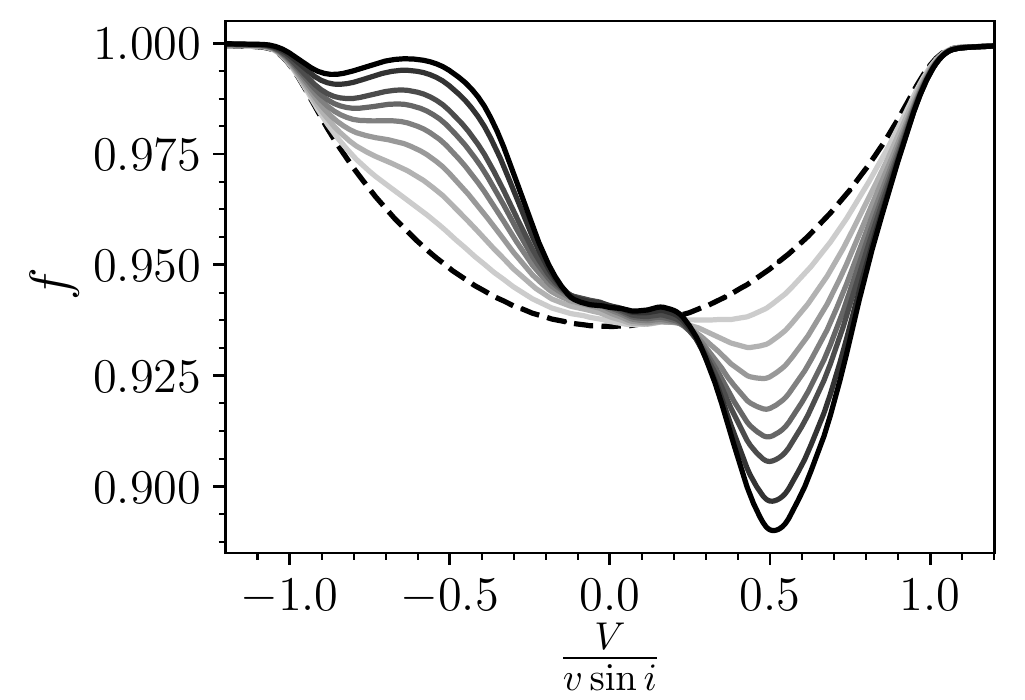}
\caption{The dependence of the distortions on the maximum optical thickness of the cloud shown in Fig.\,\ref{fig:cloud}. The dashed line is for $\tau_{\rm max}=0$, the black solid line is for $\tau_{\rm max}=5,$ the grey lines are for the intermediate cases $\tau_{\rm max}=0.5, 1, 1.5, 2, 2.5, 3, 4.$ }
\label{fig:cloud2a}
\end{center}
\end{figure}

For various lines, the obscuring cloud has the same shape; this explains why all lines show the same distortions. Some wavelength dependence should arise due to selective extinction in the clouds, i.e. the cloud's optical thickness depends on the wavelength. Line profiles for the same cloud, but for various maximum optical thicknesses $\tau_{\rm max},$ are shown in Fig.\,\ref{fig:cloud2a}.

\begin{figure}
\begin{center}
\includegraphics[scale=0.6]{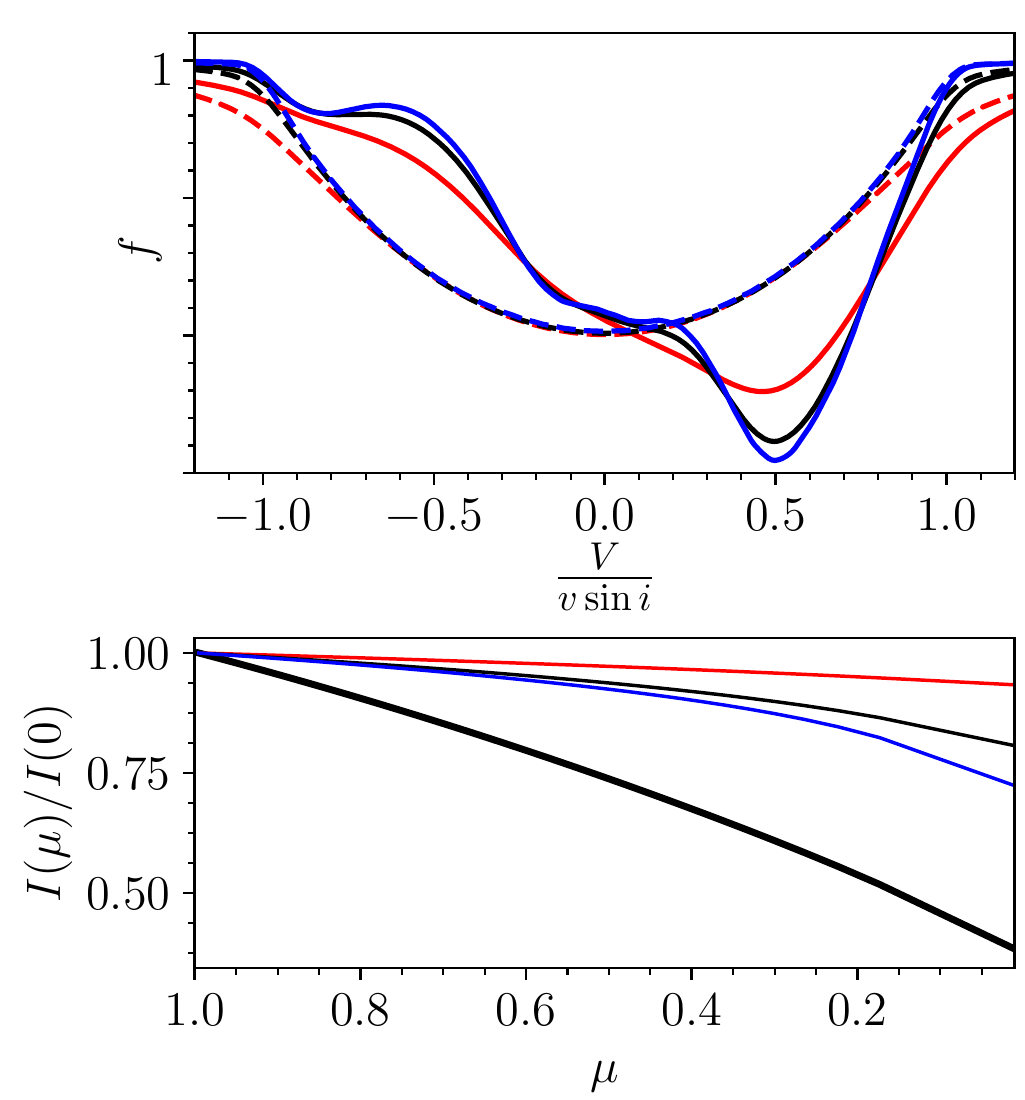}
\caption{ The dependence of the distortions on line strength. All else being equal, the oscillator strengths for the red, black, and blue lines are in the ratios of 10:1:0.1. The cloud shape and parameters are the same as in Fig.\,\ref{fig:cloud}. Profiles are scaled to the equal central depth in the out-of-eclipse state. Corresponding limb-darkening laws in the line centre are shown in the bottom panel. The thick black line is for the continuum limb-darkening.}
\label{fig:cloud2b}
\end{center}
\end{figure}

\begin{figure}
\begin{center}
\includegraphics[scale=0.6]{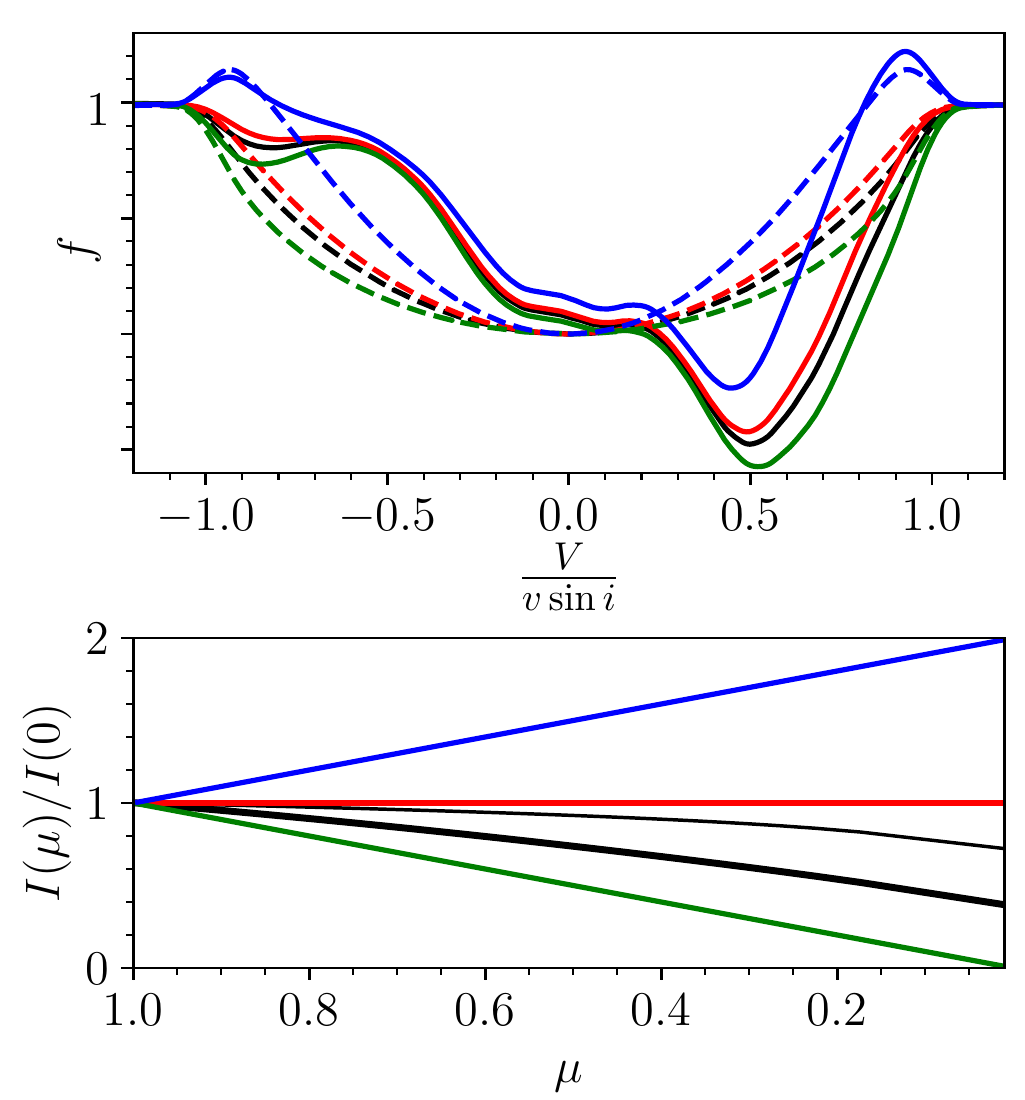}
\caption{The dependence of the distortions on the limb-darkening law. The cloud shape and parameters are the same as in Fig.\,\ref{fig:cloud}.  Corresponding limb-darkening laws in the line centre are shown in the bottom panel. The thick black line is for the continuum limb-darkening, which was the same in all cases. Thin black line is for the modelled $I(\mu)$; red, green, and blue lines correspond to linear limb-darkening with $\varepsilon=0,1,-1.$ }
\label{fig:limbDark}
\end{center}
\end{figure}

Another mechanism, which makes the distortions different, is variations in the limb darkening $I(\mu).$ In the LTE approximation, $I(\mu)$ reflects the thermal structure of the atmosphere, and generally speaking, can be different for different lines. One can expect that lines formed in the upper atmosphere will have somewhat different distortions than the weak lines formed in deeper layers (see Fig.\,\ref{fig:cloud2b}). Departures from LTE can introduce more diversity into the limb darkening, and enhance individual differences in the line distortions; in some cases, non-LTE effects can also lead to appearance of the emission lines \citep{Alexeeva2016, Sitnova2018, Mashonkina2020}. To illustrate how the diversity in the line limb-darkening can impact the distortions, we considered a simplest case of linear limb-darkening $1-\varepsilon+\varepsilon\mu$ with $\varepsilon=0,1,-1;$ the corresponding profiles are shown in Fig.\,\ref{fig:limbDark}.

\section{Mapping of the clouds}

The observed distortions of the line profiles allow us to study the circumstellar clouds by using principles of the well-known Doppler imaging technique:

1) the stellar rotation allows us to study the clouds optical thickness along the $x$-axis (orthogonal to the projection of the rotation axis): $\tau = \tau(x).$

2) variations in the limb darkening of various lines allow us to study the structure $\tau = \tau(r)$ along the radius of the stellar disc: $r^{2} = x^{2} + y^{2}.$

As well as in the case of the Doppler imaging, a series of spectral observations is needed at moments when the same cloud covers different parts of the stellar disc. Since we do not have such observations, we will restrict ourselves to a simpler approach which, nevertheless, can give some quantitative information about the degree of inhomogeneity of the dust clouds.

\subsection{The 1D mapping}

Each radial velocity in the rotational profile corresponds to a set of points on the stellar surface, which contribute to absorption in the particular spectral line. For a rigid stellar rotation, such points form straight lines parallel to the projection of the rotation axis ($y$). We will call these the {\it lines of equal radial velocity} (LERV).
Due to a limited resolution of the observed spectra and an intrinsic line width, the LERVs can be represented as finite width {\it strips} (SERV), in limits of which variations of the radial velocity are less than the width of the instrumental/intrinsic profile.

We will describe obscuration of each SERV with a mean optical thickness $\overline{\tau}$, i.e. we assume that within each SERV, the obscuration is homogeneous. In this case, the stellar spectrum $f_j$, normalized to the continuum, can be calculated as:
\begin{equation}\label{eq:1}
f_{j} = \frac{\sum I_{ij}         s_{i} e^{-\overline{\tau}_{i}a_j}}
             {\sum I^{\rm c}_{ij} s_{i} e^{-\overline{\tau}_{i}a_j}}, 
\end{equation}
where $I_{ij}$ is an average intensity at wavelength $\lambda_j$ of the $i$-th SERV with an area $s_{i}$, covered by a cloud with $\overline{\tau}_{i}$ at some reference wavelength (conventionally at the $V$-band); $a_j$ is an extinction law, normalized to unity at the reference wavelength; $I^{\rm c}_{ij}$ is an average continuum intensity. The intensity averaged over the SERV is:
\begin{equation}\label{eq:2}
I_{ij} = \frac{\int I[(\lambda_j(1 + V/c),\mu] ds_i}{s_i}.
\end{equation}
Here we account for the Doppler shift due to stellar rotation, when each point has a radial velocity $V.$ If we have at least one spectral line profile, i.e. if we know the values of $f_{j},$ we can solve equation~\ref{eq:1} with respect to $\overline{\tau}_{i}.$ 

As we have mentioned above for equation\,\ref{eq:0}, equation \ref{eq:1} also can be rewritten in a linear form: 
\begin{equation}\label{eq:1a}
f_j=\sum u_{ij}w_{i}, 
\end{equation}
with unknown weighting coefficients $w_{i},$ related to $\tau_i$ as:
\begin{equation}\label{eq:1a1}
w_{i} = \frac{I^{\rm c}_{i} s_{i} e^{-\overline{\tau}_{i}a}}
             {\sum I^{\rm c}_{i} s_{i} e^{-\overline{\tau}_{i}a}}. 
\end{equation}
Here we assume that, in the narrow wavelength range, the extinction law $a_j$ and continuum intensity $I^{\rm c}_{ij}$ do not depend on the wavelength $\lambda_j$. Each row of the matrix $u_{ij}=I_{ij}/I^{\rm c}_{ij}$ is linearly independent on the others due to Doppler shifts, therefore equation\,\ref{eq:1a} has a unique solution in terms of the weights $w_i$. However, we will solve the problem in the form of equation\,\ref{eq:1} because, in the form of equation\,\ref{eq:1a}, a negative $w_i$ can appear due to noise in the observations.  Unlike $w,$  the resulting solution in terms of $\tau$ is ambiguous: if $\overline{\tau}(x)$ is a solution, then $\overline{\tau}(x)+const$ is also a solution, because the numerator and denominator in equation~\ref{eq:1} (as well as in equation~\ref{eq:1a1}) can be multiplied by an arbitrary factor without affecting the result. This reflects the simple fact that a uniform extinction across the disc does not change the shape of the profiles. To eliminate this ambiguity, we will subtract from the obtained solution the smallest optical thickness which this solution contains. Equation \ref{eq:1} is solved by multidimensional optimization with Tikhonov's regularization, which also looks more natural expressed in $\tau,$ rather than in $w.$ The regularization parameter was chosen so that the resulting smoothing of the spectral line profiles would not noticeably exceed the spectral resolution.

\begin{figure*}
\begin{center}
\includegraphics[scale=0.57]{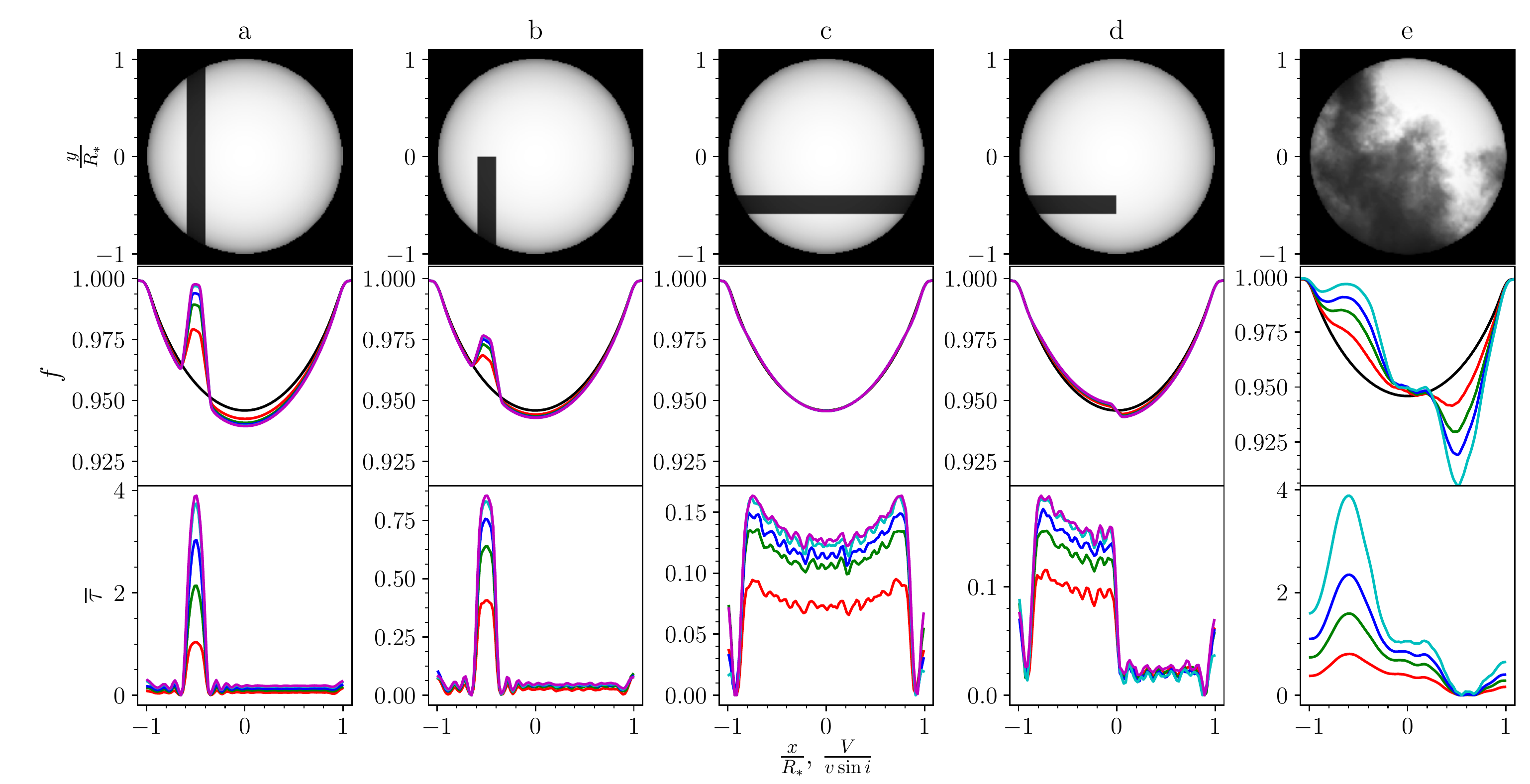}
\caption{The line profiles (in the middle row) for various geometries of the obscuring cloud (the upper row), and results of the one-dimensional mapping (the bottom row).
The simulations were performed for several values of $\tau_0$ ($\tau_{\rm max}$ in the case e): 1 (red), 2 (green), 3 (blue), 5 (cyan), and 100 (magenta, for cases a--d).}
\label{fig:test_maps}
\end{center}
\end{figure*}

To test the method described above, we first considered the simplest case, in which a homogeneous dust strip with $\tau_0$ covers the stellar disc along the $y$-axis, as shown in Fig.\,\ref{fig:test_maps}a. In this case, a bump appears in the line profiles, the height of which depends on $\tau_0.$ Solving the inverse problem allows us to recover the value of $\tau_0$ and the position of the strip on the stellar disc. In the case of partial SERV obscuration, the recovered value of $\overline{\tau}$ is smaller than $\tau_0$ because the relation $e^{-\overline{\tau}}=1-f+fe^{-\tau_0},$ where $f$ is the obscured fraction of the SERV, must hold (see Fig.\,\ref{fig:test_maps}b). The distortions will be less pronounced for clouds aligned along the $x$-axis (see Fig.\,\ref{fig:test_maps}c--d).
Extinction in the strip above a certain limit, which depends on cloud geometry, does not affect the spectrum because the contribution from the eclipsed area becomes negligible so its changes are insignificant. For this reason, strips with $\tau_0=5$ and $\tau_0=100$ lead to the same results in Fig.\,\ref{fig:test_maps}. 
Finally, for a cloud of complex shape (see Fig.\,\ref{fig:test_maps}e), it is also possible to estimate the characteristic gradient of the optical thickness.
The maximum difference of the optical thicknesses in the cloud shown in Fig.\,\ref{fig:test_maps}e is $\tau_{\rm max},$ while the restored value is $\approx0.8\tau_{\rm max}.$

The above examples demonstrate that the reconstructed  distributions $\overline{\tau}(x)$ are smoother than the original  distributions $\tau$ in the plane of the sky, but allow us to adequately estimate the extinction gradient in the direction perpendicular to the rotation axis's projection onto the plane of the sky.

\subsection{Impact of inaccuracies in line and stellar parameters}\label{sect:err}
\begin{figure*}
\begin{center}
\includegraphics[scale=0.57]{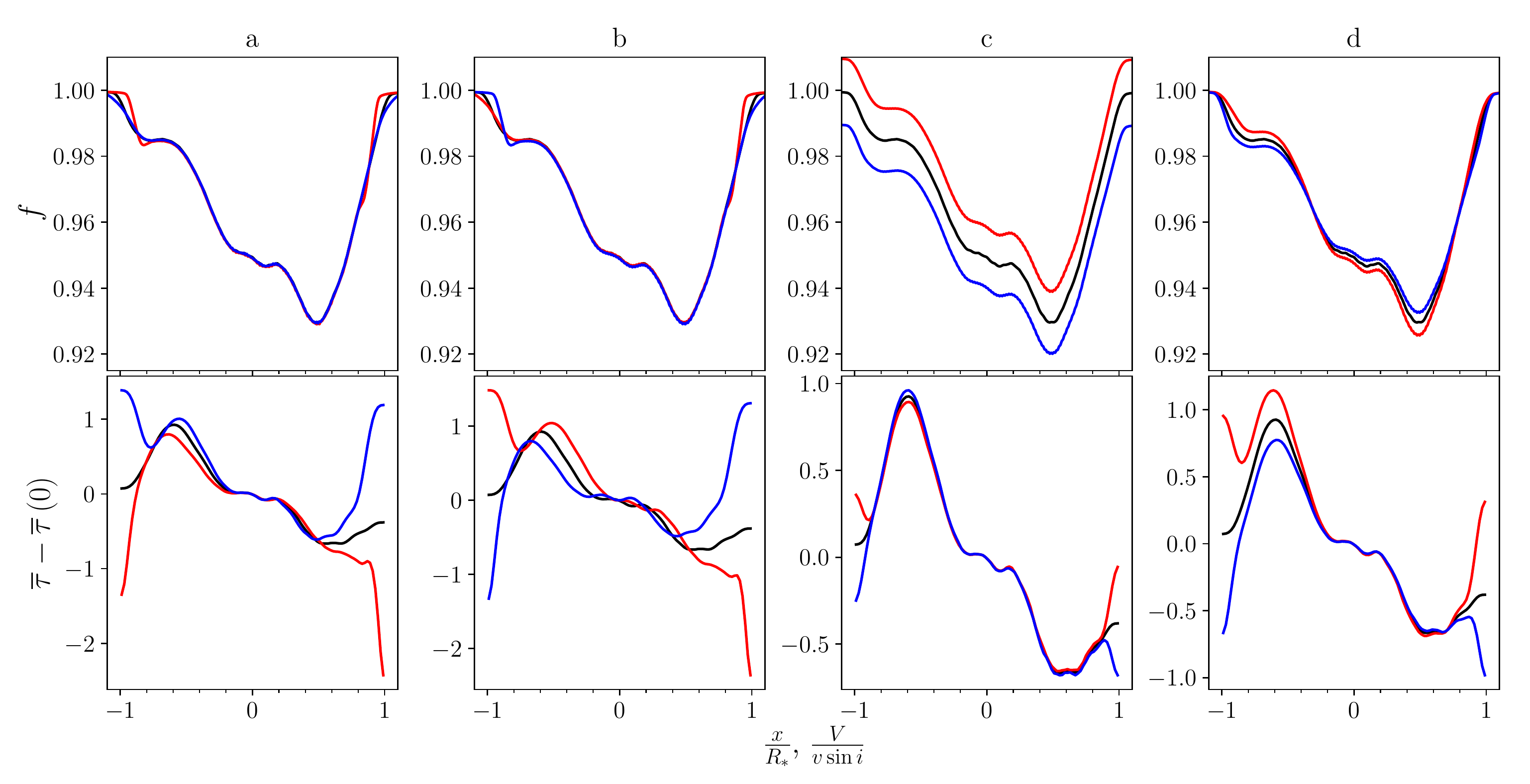}
\caption{The impact of inaccuracies on the results of 1D mapping: 
the reconstructed line profile (upper row) and the one-dimensional map $\overline{\tau}(x)$ (bottom row). The black lines are for the mapping results with the exact parameters and given for reference. The red and blue lines correspond to the results with an inaccuracy introduced into one parameter at a times:
in the rotational velocity $\frac{\Delta v \sin i}{v\sin i}=\pm 0.10$ (a),
in the radial velocity $\frac{\Delta V_{\rm r}}{V_{\rm r}}=\pm 0.10$ (b),  in the continuum level $\pm 0.01$ (c), and in the line depth $\frac{\Delta d }{d}=\pm 0.1$ (d).} 
\label{fig:test_err}
\end{center}
\end{figure*}

Reconstructing the optical thickness across the cloud requires a good model which describes the observed spectrum in the out-of-eclipse state.
We require precise values for $V_{\rm r},$ $v\sin i,$ atmospheric parameters, elemental abundances, and line-broadening parameters, in addition to the observed spectrum, accurately normalized to the continuum level. Inaccuracies in these parameters can lead to artefacts in the solution $\overline{\tau}(x).$ 
To give an impression about arising artificial features, we simulated an  `observed' spectrum, using the same cloud as in Fig.\,\ref{fig:cloud} with $\tau_{\rm max}=2,$ and then reconstructed $\overline{\tau}(x),$ varying one of the parameters (see Fig.\,\ref{fig:test_err}).

If we set $v\sin i$ a little larger than the true value, then an abnormal extinction arises at the edges of $\overline{\tau}(x),$ which compensates for a too broad line profile assumed in the modelling. The opposite situation, when $v\sin i$ is set to below its true value, leads to a strong drop of $\overline{\tau}(x)$ at the edges, and to a poorer fit. A wrong value of $V_{\rm r}$ produces artificial features at the edges of $\overline{\tau}(x)$ and leads to a poorer fit of one wing. The combined effect of inaccuracies in both $V_{\rm r}$ and $v\sin i$ can lead to asymmetric features at the edges of $\overline{\tau}(x).$ 

It can be seen from Fig.\,\ref{fig:test_err} that inaccuracies in $V_{\rm r}$ and $v\sin i$ almost do not change the extinction map in its central part, but affect only the edges, since the differences in the profiles with such inaccuracies are most pronounced on the steep wings of the line profile. A similar behaviour is observed, when we disturb the continuum normalization, but such disturbances either are clearly seen, when we compare the spectra, or do not produce strong artefacts in the map.

Artefacts due to inaccuracies in the line broadening parameters should be similar to those due to inaccuracies in $v \sin i$ and the continuum level. Uncertainties in the atmospheric parameters and the elemental abundances can be reduced to inaccuracies in the line depths and corresponding artifacts are also significant only at edges of the map $\overline{\tau}(x)$ -- see Fig.\,\ref{fig:test_err}.

One can conclude that the indirect signatures of significant uncertainties in the parameters are the steep change $\overline{\tau}(x)$ at the edges and poorly fitted line profiles. These signatures should be stable over a series of observations, if inaccuracies in the parameters are consistent throughout. The most reliable information on the extinction gradients is contained in the central part of the profile, which is less sensitive to inaccuracies in the line and stellar parameters.

\subsection{The 1D mapping on the real spectra}\label{sect:map}

\begin{figure*}
\begin{center}
\includegraphics[scale=0.56]{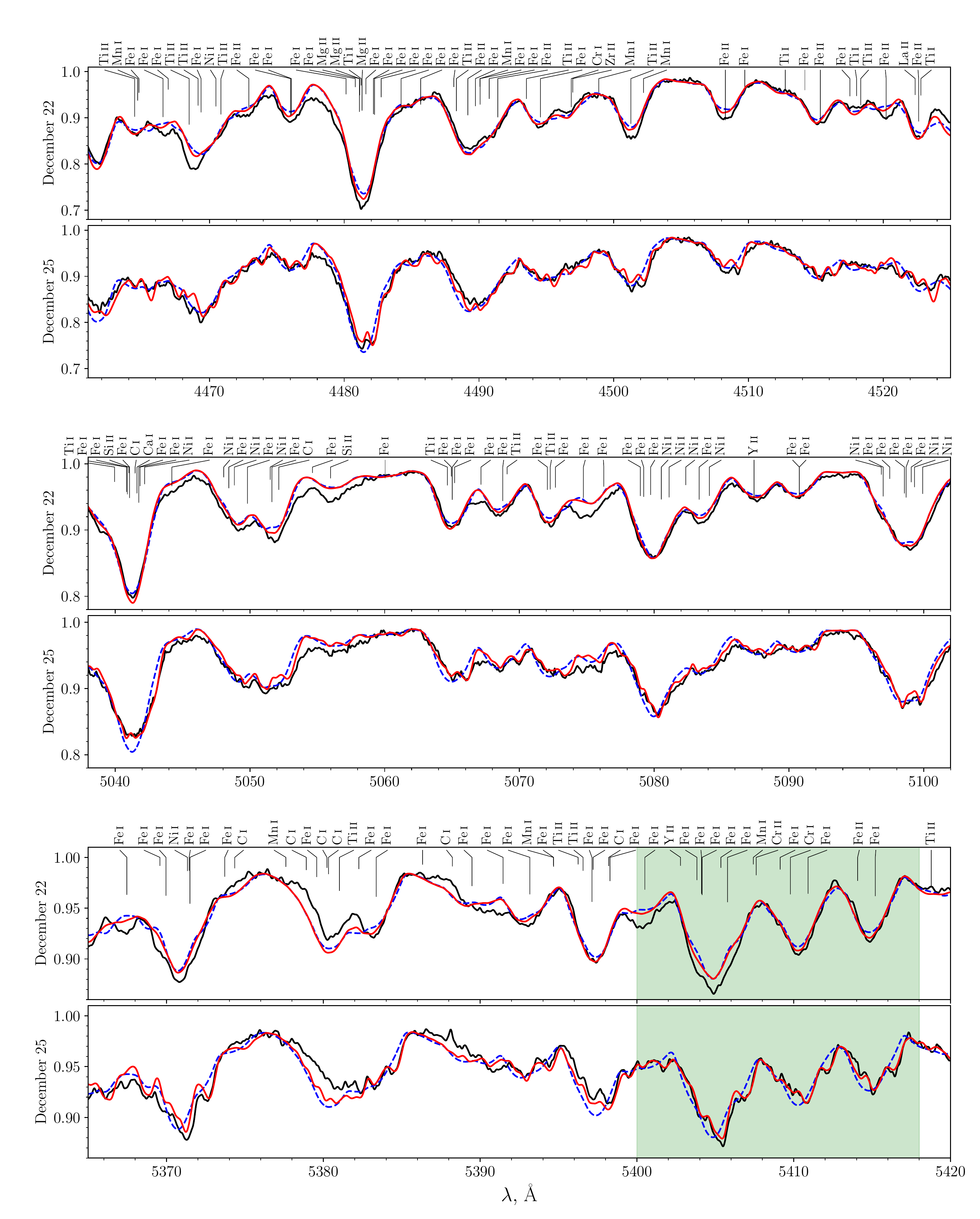}
\caption{The observed spectra (black lines) of CQ Tau for December 22 (the upper parts) and December 25 (the lower parts). The dashed blue line is the out-of-eclipse synthetic spectrum of CQ Tau. The red line is the synthetic spectra during the eclipses with the recovered 1D distributions $\overline{\tau}(x),$ which are shown in Fig.\,\ref{fig:obs_err}. The shaded area is a region which was used to reconstruct $\overline{\tau}(x).$}
\label{fig:obs245}
\end{center}
\end{figure*}
\addtocounter{figure}{-1}
\begin{figure*}
\begin{center}
\includegraphics[scale=0.56]{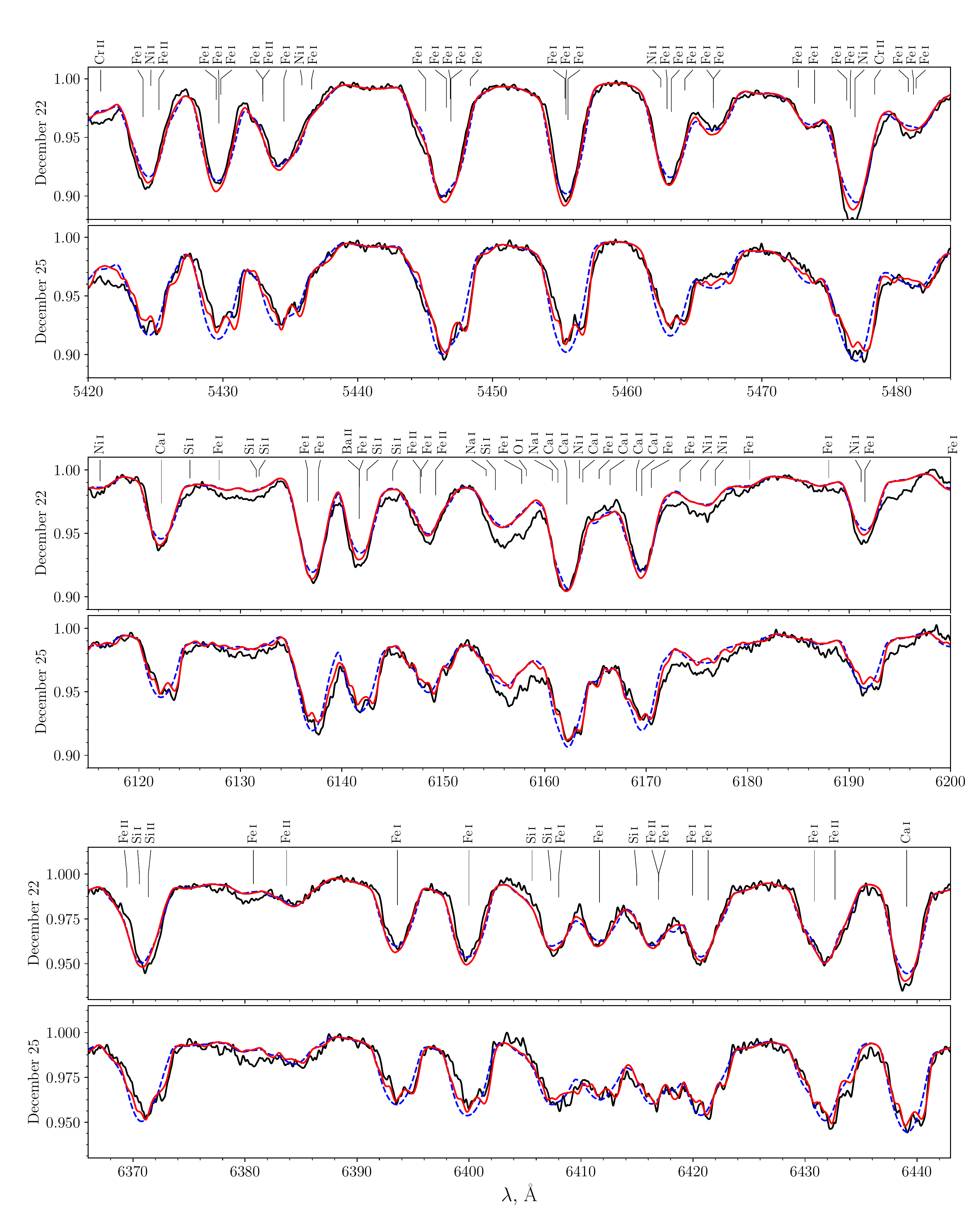}
\caption{Continued}
\label{fig:obs245_2}
\end{center}
\end{figure*}

\begin{figure}
\begin{center}
\includegraphics[scale=0.6]{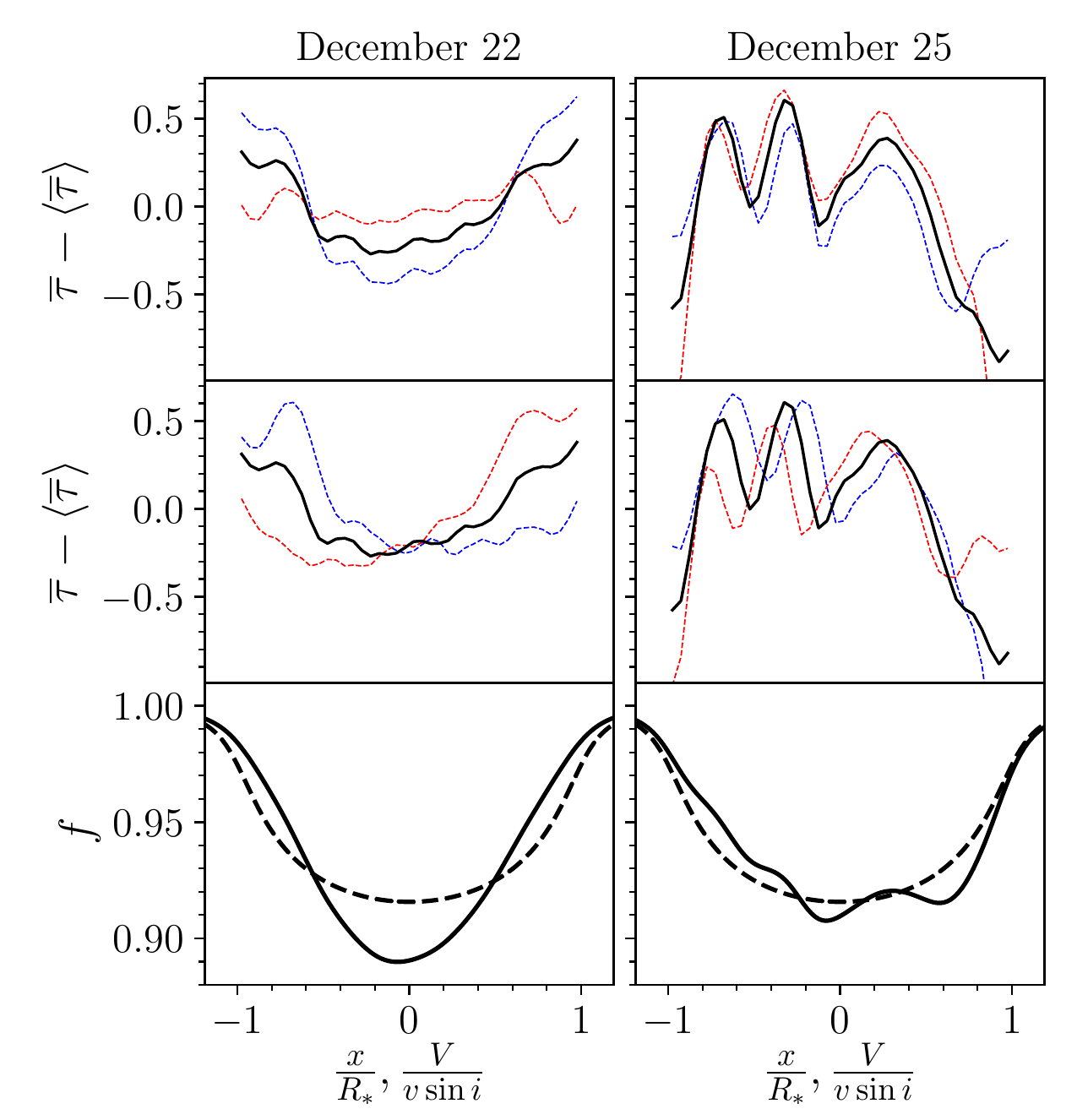}
\caption{The 1D map $\overline{\tau}(x)$ reconstructed over the spectral region 5400 -- 5418{\AA} (the shaded area in Fig.\,\ref{fig:obs245}). The dashed lines show the map distorted by varying $v\sin i$ (the upper panel) or $V_{\rm r}$ (the middle panel)  by $+10$\,km\,s$^{-1}$ (the blue line) and $-10$\,km\,s$^{-1}$ (the red line). The mean value $\langle\overline{\tau}\rangle$ was removed for ease of comparison. The bottom panels are for the line profiles, the dashed lines are for the out-of-eclipse spectrum, the solid lines correspond to the obscured star with $\overline{\tau}(x),$ shown by black lines in the panels above.}
\label{fig:obs_err}
\end{center}
\end{figure}

We apply the 1D mapping method described above to the archival spectra of CQ Tau. To simulate instrumental broadening, the theoretical spectra were additionally smoothed with a Voigt profile, which reproduces the far wings of the observed profiles better than a simple Gaussian broadening. We show, in Fig.\,\ref{fig:obs245}, 6 spectral regions from blue to red for the spectra of December 22, when the lines looked less distorted, and also of December 25, when the strongest distortions were observed. Equation\,\ref{eq:1} can be solved if we have at least one profile that allows us to use a small spectral region to reconstruct $\overline{\tau}(x),$ while the remaining spectrum may be used to validate our explanation of the distortions. We chose a spectral region with a few \ion{Fe}{i} lines (shadowed in Fig.\,\ref{fig:obs245}). The corresponding solutions $\overline{\tau}(x)$ are shown in Fig.\,\ref{fig:obs_err} for both nights. To calculate the synthetic spectra over a wide spectral range, we assume the standard extinction law \citep{Car89}. For December 22, the amplitude of variations in $\overline{\tau}$ is smaller, and significant absorption appears only at edges, and therefore depends on the assumed values for $v \sin i$ and $V_{\rm r}$ (see Section\,\ref{sect:err}). Since we do not have out-of-eclipse spectra, the applied values of $v \sin i$ and $V_{\rm r}$ can be uncertain. To check how these uncertainties can impact our results, we vary $v \sin i$ and $V_{\rm r}$ by $\pm10$\,km\,s$^{-1};$ the corresponding $\overline{\tau}(x)$ are shown with dashed lines. Unlike on December 22, absorption peaks in $\overline{\tau}$ for December 25 appear in the central part. In this case, inaccuracies in $v \sin i$ and $V_{\rm r}$ are less important, which allows us to estimate the characteristic gradient and size of eclipsing clouds even without exact values of $v \sin i$ and $V_{\rm r}.$
 
We see three peaks in the map at $x/R_*\approx-0.70,$ $-0.30,$ 0.27. The first peak ($-0.70$) makes the left wing of the profile less steep than the right one: see the line profile in the bottom panel of Fig.\,\ref{fig:obs_err}. There are also other mechanisms which can produce asymmetric wings, for example, scattering on the moving dust clouds. This peak is also sensitive to the assumed radial velocity of the star, so we cannot be sure of its parameters. However,  the peaks at $x\approx -0.30R_*$ and $0.27R_*$  are quite reliable, and are responsible for the appearance of the jagged profiles. The most pronounced bump in the profile is produced by the peak at $0.27R_*,$ which has an amplitude $\Delta A_{V}\sim\Delta\overline{\tau}\sim 1$ and a linear size $\approx0.6R_*.$  It means the existence of spatially small but optically thick dusty clouds on the line of sight. Since the 1D mapping underestimates the extinction gradients, the real inhomogeneity could be much larger.

Because any homogeneous extinction can be added to the obtained solution $\overline{\tau},$ we can estimate only a lower limit on a brightness drop $\Delta V,$ which is 0.4 and 1.0 magnitude for December 22 and 25, correspondingly. The observed drop is $1.5-2$ magnitudes, therefore a residual $A_{V}\approx 1$ is related to a flat part of the circumstellar extinction.

Despite the fact that only three lines were used to reconstruct the map, it reproduces the distortions over a broad spectral region: narrow features in the line profiles appear at the very places where our model predicts. This confirms our interpretation of the observed distortions as a result of the inhomogeneous obscuration of the rotating stellar surface. Visible discrepancies between the theoretical and observed spectrum are mostly in the equivalent widths and may be due to uncertainties in the atmospheric parameters, non-LTE effects, or errors in the line parameters.

\section{Discussion}

CQ Tau shows accretion signatures (inverse P-Cygni profiles of \ion{He}{i} 5876{\AA} and \ion{O}{i} 7773{\AA}), and one can suppose that the variability observed in the profiles is related to the veiling by lines \citep{Dodin2012} when the stellar absorption lines are filled-in by weak emission lines, which can appear in various parts of the profiles depending on the geometry and orientation of the accretion hotspot. However, these emissions should reduce the total equivalent widths of the lines, which is not observed. Moreover, observations as well as simulations show that these emissions are stronger for stronger lines, which is also not observed in our case. Therefore, we conclude that the accretion region is not responsible for the line distortions; moreover, the absence of a noticeable veiling indicates that the accretion contribution is negligible and visible only in the strongest lines: H~$\alpha$, \ion{He}{i} 5876\,{\AA} and \ion{O}{i} 7773\,{\AA}. In principle, the accretion process can produce an extra UV radiation field, which can cause Fe overionization and hence enhance \ion{Fe}{ii} lines (see Section\,\ref{sect:prop}).

Cold spots on the stellar surface can also distort the line profiles in a manner similar to what has been observed. However, CQ Tau does not have a convective envelope (see fig.\,1 in \citealt{Vil19}), and cold spots on its surface are not expected. Moreover, such spots should produce a noticeable regular photometric variability, which has not been detected.

One can conjecture that the distortions arise due to some circumstellar gas, in which case the equivalent widths should increase; this also is not observed. In addition, the distortions of the profiles would appear first of all in the strongest lines: in the case of a cold gas, in \ion{Na}{i}, \ion{K}{i} lines, or else in \ion{H}{i}, \ion{He}{i} lines, in the case of hotter gas or accretion flows \citep{Dodin18}. These absorption features are indeed present in the spectrum of CQ Tau, but they have a completely different shape which does not resemble the distortions in all of the other absorption lines. 

Because the distortions do not depend on the physical parameters of the lines, it is natural to suppose that they are induced by geometrical effects: both or either of inhomogeneous obscuration, and scattering by moving dust clouds \citep{GT12}.

Light scattering on moving clouds can modify line profiles; these changes depend on the geometry and kinematics of the scattering bodies. Such scattering typically leads to profile smoothing, rather than to the appearance of sharp, narrow features. One can suppose a very special geometry, in which scattering clouds `see' the star pole-on, while the observer looks in the equatorial plane. In this case, the lines in the scattered spectrum will be broadened less by the stellar rotation than they are in direct light. However, such an explanation looks too artificial because it requires: 
1) a simultaneous presence of a few clouds, moving with various velocities, but just within $v \sin i$; and 2) the contribution of scattered light from these clouds must be comparable to the contribution of direct light from the star, in order for the distortions to be noticeable.
Simultaneously satisfying both of these requirements seems unlikely, however, noticeable contribution of scattered light must be present because CQ Tau's $V$-band magnitude was $10.6-11.$ This magnitude corresponds to the colour turn on the colour--magnitude diagram \citep{Shakh05}, meaning that about half of the flux is provided by scattered light.
The peak at $x/R_*=-0.7$ in $\overline{\tau}(x)$ can probably be explained by scattered light, however, to do this, special modelling is needed, which is outside the scope of this study. Nevertheless, we can understand qualitatively how scattering can change our results. Since the profile is smoothed by the scattering on an extended environment (a rotating disc or an expanding envelope), a spectrum with smoother profiles is added to the jagged spectrum of the direct stellar light, reducing the amplitude of the distortions. This reduction leads to a corresponding decrease in the extinction gradients. Thus, the map $\overline{\tau}$ obtained without taking scattered light into account gives a lower limit on the extinction gradients, while accounting for the scattering should lead to a sharper map.

Since we have not detected variability over 2 hours of observations, the cloud's crossing time $t_{\rm c}\gtrsim0.5$ day, but the spectra on December 22 and 25 are completely different so, therefore, $t_{\rm c}\lesssim3$ days. For the stellar parameters from \citet{Vil19}, we can estimate that, for Keplerian velocities, such crossing times correspond to distances between $0.9-33$ a.u., i.e. well outside the sublimation radius ($\sim 0.4$ a.u.).

\section{Conclusions}
Using CQ Tau as an example, we have shown that the circumstellar dust clouds are inhomogeneous on scales of the stellar radius that leads to variable distortions in the line profiles. A distinctive feature of the line distortions, caused by the inhomogeneous obscuration, is the approximate conservation of the equivalent widths. In the case of another known source of profile variability in young stars -- the accretion hotspots -- the line equivalent width is reduced, moreover this effect is stronger for stronger lines \citep{Dodin2012, Dodin18, Rei18}. Cold spots on the stellar surface also almost do not change the equivalent width because the absolutely dark spot is equivalent to an absolutely opaque cloud, however, the time behaviour of the distortions should differ between these two cases.

The spectral variability of CQ Tau shows us that the star cannot be considered as a point source when we deal with circumstellar dust. This fact significantly complicates the interpretation of various kinds of observations: the line distortions prevent an accurate determination of the stellar parameters, first of all, $V_{\rm r}$ and $v \sin i$; an inhomogeneous obscuration makes the apparent extinction law greyer than the intrinsic one \citep{Natta1984}; an inhomogeneous obscuration can change the relative contribution of hot/cold spots on the stellar surface, for example, it can modify the apparent accretion indicators, if they are measured relative to the stellar spectrum (e.g., equivalent width, veiling), since the obscuration of the accretion region and the stellar surface can be different and variable. Perhaps this explains why there is no correlation between accretion tracers and stellar brightness in RW Aur\,A (see fig.\,11 in \citealt{Petrov01}).

On the other hand, the line distortions offer the possibility of studying the dust clouds on scales inaccessible by direct observation. The estimates of the size and amplitude of dust clumps, given above, were obtained from merely one archival spectrum. We encourage special spectral observations, which should cover various phases of a dust clump transit over the stellar disc. Such observations will allow the velocity of the clouds to be measured relative to the rotational axis of the star, and to clarify what obscures the star: whether a clumped dusty wind or a warped disc.

The clumpy structure of the dust environment probably also takes place for other types of stars. However, in many cases, it will be difficult to study by the method proposed here, due to a slower rotation and other sources of spectral and photometric variability (such as accretion spots or cold spots in the case of T Tauri stars).

\section*{Acknowledgements}
This work is supported by the Russian Science Foundation under grant 20-72-10011.

\section*{Data availability}

The data underlying this article were accessed from Canadian Astronomy Data Centre \url{http://www.cadc-ccda.hia-iha.nrc-cnrc.gc.ca}. The derived data generated in this research will be shared on reasonable request to the corresponding author.




\bibliographystyle{mnras}
\bibliography{uxori} 

\begin{thebibliography}{}
\makeatletter
\relax
\def\mn@urlcharsother{\let\do\@makeother \do\$\do\&\do\#\do\^\do\_\do\%\do\~}
\def\mn@doi{\begingroup\mn@urlcharsother \@ifnextchar [ {\mn@doi@}
  {\mn@doi@[]}}
\def\mn@doi@[#1]#2{\def\@tempa{#1}\ifx\@tempa\@empty \href
  {http://dx.doi.org/#2} {doi:#2}\else \href {http://dx.doi.org/#2} {#1}\fi
  \endgroup}
\def\mn@eprint#1#2{\mn@eprint@#1:#2::\@nil}
\def\mn@eprint@arXiv#1{\href {http://arxiv.org/abs/#1} {{\tt arXiv:#1}}}
\def\mn@eprint@dblp#1{\href {http://dblp.uni-trier.de/rec/bibtex/#1.xml}
  {dblp:#1}}
\def\mn@eprint@#1:#2:#3:#4\@nil{\def\@tempa {#1}\def\@tempb {#2}\def\@tempc
  {#3}\ifx \@tempc \@empty \let \@tempc \@tempb \let \@tempb \@tempa \fi \ifx
  \@tempb \@empty \def\@tempb {arXiv}\fi \@ifundefined
  {mn@eprint@\@tempb}{\@tempb:\@tempc}{\expandafter \expandafter \csname
  mn@eprint@\@tempb\endcsname \expandafter{\@tempc}}}

\bibitem[\protect\citeauthoryear{{Alecian} et~al.,}{{Alecian}
  et~al.}{2013}]{Al13}
{Alecian} E.,  et~al., 2013, \mn@doi [\mnras] {10.1093/mnras/sts383}, \href
  {https://ui.adsabs.harvard.edu/abs/2013MNRAS.429.1001A} {429, 1001}

\bibitem[\protect\citeauthoryear{{Alexeeva}, {Ryabchikova}  \&
  {Mashonkina}}{{Alexeeva} et~al.}{2016}]{Alexeeva2016}
{Alexeeva} S.~A.,  {Ryabchikova} T.~A.,   {Mashonkina} L.~I.,  2016, \mn@doi
  [\mnras] {10.1093/mnras/stw1635}, \href
  {https://ui.adsabs.harvard.edu/abs/2016MNRAS.462.1123A} {462, 1123}

\bibitem[\protect\citeauthoryear{{Barsunova}, {Grinin}, {Sergeev}, {Semenov}
  \& {Shugarov}}{{Barsunova} et~al.}{2015}]{B15}
{Barsunova} O.~Y.,  {Grinin} V.~P.,  {Sergeev} S.~G.,  {Semenov} A.~O.,
  {Shugarov} S.~Y.,  2015, \mn@doi [Astrophysics] {10.1007/s10511-015-9375-8},
  \href {https://ui.adsabs.harvard.edu/abs/2015Ap.....58..193B} {58, 193}

\bibitem[\protect\citeauthoryear{{Bergemann}, {Lind}, {Collet}, {Magic}  \&
  {Asplund}}{{Bergemann} et~al.}{2012}]{Ber12}
{Bergemann} M.,  {Lind} K.,  {Collet} R.,  {Magic} Z.,   {Asplund} M.,  2012,
  \mn@doi [\mnras] {10.1111/j.1365-2966.2012.21687.x}, \href
  {https://ui.adsabs.harvard.edu/abs/2012MNRAS.427...27B} {427, 27}

\bibitem[\protect\citeauthoryear{{Bibo}, {The}  \& {Dawanas}}{{Bibo}
  et~al.}{1992}]{B92}
{Bibo} E.~A.,  {The} P.~S.,   {Dawanas} D.~N.,  1992, \aap, \href
  {https://ui.adsabs.harvard.edu/abs/1992A&A...260..293B} {260, 293}

\bibitem[\protect\citeauthoryear{{Bouvier} et~al.,}{{Bouvier}
  et~al.}{1999}]{BCAC99}
{Bouvier} J.,  et~al., 1999, \aap, \href
  {https://ui.adsabs.harvard.edu/abs/1999A&A...349..619B} {349, 619}

\bibitem[\protect\citeauthoryear{{Cardelli}, {Clayton}  \& {Mathis}}{{Cardelli}
  et~al.}{1989}]{Car89}
{Cardelli} J.~A.,  {Clayton} G.~C.,   {Mathis} J.~S.,  1989, \mn@doi [\apj]
  {10.1086/167900}, \href
  {https://ui.adsabs.harvard.edu/abs/1989ApJ...345..245C} {345, 245}

\bibitem[\protect\citeauthoryear{{Cody} et~al.,}{{Cody} et~al.}{2014}]{Cody14}
{Cody} A.~M.,  et~al., 2014, \mn@doi [\aj] {10.1088/0004-6256/147/4/82}, \href
  {https://ui.adsabs.harvard.edu/abs/2014AJ....147...82C} {147, 82}

\bibitem[\protect\citeauthoryear{{Cohen}, {Herbst}  \& {Williams}}{{Cohen}
  et~al.}{2003}]{CHW03}
{Cohen} R.~E.,  {Herbst} W.,   {Williams} E.~C.,  2003, \mn@doi [\apjl]
  {10.1086/379275}, \href
  {https://ui.adsabs.harvard.edu/abs/2003ApJ...596L.243C} {596, L243}

\bibitem[\protect\citeauthoryear{{Dodin}}{{Dodin}}{2018}]{Dodin18}
{Dodin} A.,  2018, \mn@doi [\mnras] {10.1093/mnras/sty038}, \href
  {https://ui.adsabs.harvard.edu/abs/2018MNRAS.475.4367D} {475, 4367}

\bibitem[\protect\citeauthoryear{{Dodin} \& {Lamzin}}{{Dodin} \&
  {Lamzin}}{2012}]{Dodin2012}
{Dodin} A.~V.,  {Lamzin} S.~A.,  2012, \mn@doi [Astronomy Letters]
  {10.1134/S1063773712100027}, \href
  {https://ui.adsabs.harvard.edu/abs/2012AstL...38..649D} {38, 649}

\bibitem[\protect\citeauthoryear{{Dodin} et~al.,}{{Dodin} et~al.}{2019}]{D19}
{Dodin} A.,  et~al., 2019, \mn@doi [\mnras] {10.1093/mnras/sty2988}, \href
  {https://ui.adsabs.harvard.edu/abs/2019MNRAS.482.5524D} {482, 5524}

\bibitem[\protect\citeauthoryear{{Friedemann}, {Reimann}  \&
  {Guertler}}{{Friedemann} et~al.}{1992}]{F92}
{Friedemann} C.,  {Reimann} H.~G.,   {Guertler} J.,  1992, \aap, \href
  {https://ui.adsabs.harvard.edu/abs/1992A&A...255..246F} {255, 246}

\bibitem[\protect\citeauthoryear{{Gahm}, {Petrov}, {Tambovsteva}, {Grinin},
  {Stempels}  \& {Walter}}{{Gahm} et~al.}{2018}]{Gahm18}
{Gahm} G.~F.,  {Petrov} P.~P.,  {Tambovsteva} L.~V.,  {Grinin} V.~P.,
  {Stempels} H.~C.,   {Walter} F.~M.,  2018, \mn@doi [\aap]
  {10.1051/0004-6361/201832891}, \href
  {https://ui.adsabs.harvard.edu/abs/2018A&A...614A.117G} {614, A117}

\bibitem[\protect\citeauthoryear{{Grinin} \& {Potravnov}}{{Grinin} \&
  {Potravnov}}{2013}]{GP13}
{Grinin} V.~P.,  {Potravnov} I.~S.,  2013, \mn@doi [Astrophysics]
  {10.1007/s10511-013-9262-0}, \href
  {https://ui.adsabs.harvard.edu/abs/2013Ap.....56....1G} {56, 1}

\bibitem[\protect\citeauthoryear{{Grinin}, {Barsunova}, {Shugarov}, {Kroll}  \&
  {Sergeev}}{{Grinin} et~al.}{2008}]{Grinin2008}
{Grinin} V.~P.,  {Barsunova} O.~Y.,  {Shugarov} S.~Y.,  {Kroll} P.,   {Sergeev}
  S.~G.,  2008, \mn@doi [Astrophysics] {10.1007/s10511-008-0001-x}, \href
  {https://ui.adsabs.harvard.edu/abs/2008Ap.....51....1G} {51, 1}

\bibitem[\protect\citeauthoryear{{Grinin}, {Tambovtseva}  \&
  {Weigelt}}{{Grinin} et~al.}{2012}]{GT12}
{Grinin} V.~P.,  {Tambovtseva} L.~V.,   {Weigelt} G.,  2012, \mn@doi [\aap]
  {10.1051/0004-6361/201219768}, \href
  {https://ui.adsabs.harvard.edu/abs/2012A&A...544A..45G} {544, A45}

\bibitem[\protect\citeauthoryear{{Grinin}, {Semenov}, {Barsunova}  \&
  {Sergeev}}{{Grinin} et~al.}{2019}]{G19}
{Grinin} V.~P.,  {Semenov} A.~O.,  {Barsunova} O.~Y.,   {Sergeev} S.~G.,  2019,
  \mn@doi [Astrophysics] {10.1007/s10511-019-09562-x}, \href
  {https://ui.adsabs.harvard.edu/abs/2019Ap.....62...41G} {62, 41}

\bibitem[\protect\citeauthoryear{{G{\"u}nther} et~al.,}{{G{\"u}nther}
  et~al.}{2018}]{G18}
{G{\"u}nther} H.~M.,  et~al., 2018, \mn@doi [\aj] {10.3847/1538-3881/aac9bd},
  \href {https://ui.adsabs.harvard.edu/abs/2018AJ....156...56G} {156, 56}

\bibitem[\protect\citeauthoryear{{Gustafsson}, {Edvardsson}, {Eriksson},
  {J{\o}rgensen}, {Nordlund}  \& {Plez}}{{Gustafsson} et~al.}{2008}]{Gus08}
{Gustafsson} B.,  {Edvardsson} B.,  {Eriksson} K.,  {J{\o}rgensen} U.~G.,
  {Nordlund} {\r{A}}.,   {Plez} B.,  2008, \mn@doi [\aap]
  {10.1051/0004-6361:200809724}, \href
  {https://ui.adsabs.harvard.edu/abs/2008A&A...486..951G} {486, 951}

\bibitem[\protect\citeauthoryear{{Hern{\'a}ndez}, {Calvet}, {Brice{\~n}o},
  {Hartmann}  \& {Berlind}}{{Hern{\'a}ndez} et~al.}{2004}]{HC04}
{Hern{\'a}ndez} J.,  {Calvet} N.,  {Brice{\~n}o} C.,  {Hartmann} L.,
  {Berlind} P.,  2004, \mn@doi [\aj] {10.1086/381908}, \href
  {https://ui.adsabs.harvard.edu/abs/2004AJ....127.1682H} {127, 1682}

\bibitem[\protect\citeauthoryear{{Kausch} et~al.,}{{Kausch}
  et~al.}{2015}]{Kausch2015}
{Kausch} W.,  et~al., 2015, \mn@doi [\aap] {10.1051/0004-6361/201423909}, \href
  {https://ui.adsabs.harvard.edu/abs/2015A&A...576A..78K} {576, A78}

\bibitem[\protect\citeauthoryear{{Manset}, {Bastien}, {M{\'e}nard}, {Bertout},
  {Le van Suu}  \& {Boivin}}{{Manset} et~al.}{2009}]{MB09}
{Manset} N.,  {Bastien} P.,  {M{\'e}nard} F.,  {Bertout} C.,  {Le van Suu} A.,
   {Boivin} L.,  2009, \mn@doi [\aap] {10.1051/0004-6361/200810945}, \href
  {https://ui.adsabs.harvard.edu/abs/2009A&A...499..137M} {499, 137}

\bibitem[\protect\citeauthoryear{{Mashonkina}}{{Mashonkina}}{2020}]{Mashonkina2020}
{Mashonkina} L.,  2020, \mn@doi [\mnras] {10.1093/mnras/staa653}, \href
  {https://ui.adsabs.harvard.edu/abs/2020MNRAS.493.6095M} {493, 6095}

\bibitem[\protect\citeauthoryear{{McGinnis} et~al.,}{{McGinnis}
  et~al.}{2015}]{MA15}
{McGinnis} P.~T.,  et~al., 2015, \mn@doi [\aap] {10.1051/0004-6361/201425475},
  \href {https://ui.adsabs.harvard.edu/abs/2015A&A...577A..11M} {577, A11}

\bibitem[\protect\citeauthoryear{{Natta} \& {Panagia}}{{Natta} \&
  {Panagia}}{1984}]{Natta1984}
{Natta} A.,  {Panagia} N.,  1984, \mn@doi [\apj] {10.1086/162681}, \href
  {https://ui.adsabs.harvard.edu/abs/1984ApJ...287..228N} {287, 228}

\bibitem[\protect\citeauthoryear{{Petrov} \& {Kozack}}{{Petrov} \&
  {Kozack}}{2007}]{PK07}
{Petrov} P.~P.,  {Kozack} B.~S.,  2007, \mn@doi [Astronomy Reports]
  {10.1134/S1063772907060091}, \href
  {https://ui.adsabs.harvard.edu/abs/2007ARep...51..500P} {51, 500}

\bibitem[\protect\citeauthoryear{{Petrov}, {Gahm}, {Gameiro}, {Duemmler},
  {Ilyin}, {Laakkonen}, {Lago}  \& {Tuominen}}{{Petrov}
  et~al.}{2001}]{Petrov01}
{Petrov} P.~P.,  {Gahm} G.~F.,  {Gameiro} J.~F.,  {Duemmler} R.,  {Ilyin}
  I.~V.,  {Laakkonen} T.,  {Lago} M.~T.~V.~T.,   {Tuominen} I.,  2001, \mn@doi
  [\aap] {10.1051/0004-6361:20010203}, \href
  {https://ui.adsabs.harvard.edu/abs/2001A&A...369..993P} {369, 993}

\bibitem[\protect\citeauthoryear{{Petrov}, {Gahm}, {Djupvik}, {Babina},
  {Artemenko}  \& {Grankin}}{{Petrov} et~al.}{2015}]{PG15}
{Petrov} P.~P.,  {Gahm} G.~F.,  {Djupvik} A.~A.,  {Babina} E.~V.,  {Artemenko}
  S.~A.,   {Grankin} K.~N.,  2015, \mn@doi [\aap]
  {10.1051/0004-6361/201525845}, \href
  {https://ui.adsabs.harvard.edu/abs/2015A&A...577A..73P} {577, A73}

\bibitem[\protect\citeauthoryear{Piskunov \& Valenti}{Piskunov \&
  Valenti}{2016}]{Pi16}
Piskunov N.,  Valenti J.~A.,  2016, \mn@doi [Astronomy \& Astrophysics]
  {10.1051/0004-6361/201629124}, 597, A16

\bibitem[\protect\citeauthoryear{{Rei}, {Petrov}  \& {Gameiro}}{{Rei}
  et~al.}{2018}]{Rei18}
{Rei} A.~C.~S.,  {Petrov} P.~P.,   {Gameiro} J.~F.,  2018, \mn@doi [\aap]
  {10.1051/0004-6361/201731444}, \href
  {https://ui.adsabs.harvard.edu/abs/2018A&A...610A..40R} {610, A40}

\bibitem[\protect\citeauthoryear{{Rodriguez} et~al.,}{{Rodriguez}
  et~al.}{2016}]{R16}
{Rodriguez} J.~E.,  et~al., 2016, \mn@doi [\apj] {10.3847/0004-637X/831/1/74},
  \href {https://ui.adsabs.harvard.edu/abs/2016ApJ...831...74R} {831, 74}

\bibitem[\protect\citeauthoryear{{Ryabchikova}, {Piskunov}, {Kurucz},
  {Stempels}, {Heiter}, {Pakhomov}  \& {Barklem}}{{Ryabchikova}
  et~al.}{2015}]{R15}
{Ryabchikova} T.,  {Piskunov} N.,  {Kurucz} R.~L.,  {Stempels} H.~C.,  {Heiter}
  U.,  {Pakhomov} Y.,   {Barklem} P.~S.,  2015, \mn@doi [\physscr]
  {10.1088/0031-8949/90/5/054005}, \href
  {https://ui.adsabs.harvard.edu/abs/2015PhyS...90e4005R} {90, 054005}

\bibitem[\protect\citeauthoryear{{Schmidt}}{{Schmidt}}{2019}]{Schmidt2019}
{Schmidt} E.~G.,  2019, \mn@doi [\apjl] {10.3847/2041-8213/ab2e77}, \href
  {https://ui.adsabs.harvard.edu/abs/2019ApJ...880L...7S} {880, L7}

\bibitem[\protect\citeauthoryear{{Schneider} et~al.,}{{Schneider}
  et~al.}{2015a}]{SchG15}
{Schneider} P.~C.,  et~al., 2015a, \mn@doi [\aap]
  {10.1051/0004-6361/201527237}, \href
  {https://ui.adsabs.harvard.edu/abs/2015A&A...584L...9S} {584, L9}

\bibitem[\protect\citeauthoryear{{Schneider}, {France}, {G{\"u}nther},
  {Herczeg}, {Robrade}, {Bouvier}, {McJunkin}  \& {Schmitt}}{{Schneider}
  et~al.}{2015b}]{SchnG15}
{Schneider} P.~C.,  {France} K.,  {G{\"u}nther} H.~M.,  {Herczeg} G.,
  {Robrade} J.,  {Bouvier} J.,  {McJunkin} M.,   {Schmitt} J.~H.~M.~M.,  2015b,
  \mn@doi [\aap] {10.1051/0004-6361/201425583}, \href
  {https://ui.adsabs.harvard.edu/abs/2015A&A...584A..51S} {584, A51}

\bibitem[\protect\citeauthoryear{{Shakhovskoj}, {Grinin}  \&
  {Rostopchina}}{{Shakhovskoj} et~al.}{2005}]{Shakh05}
{Shakhovskoj} D.~N.,  {Grinin} V.~P.,   {Rostopchina} A.~N.,  2005, \mn@doi
  [Astrophysics] {10.1007/s10511-005-0014-7}, \href
  {https://ui.adsabs.harvard.edu/abs/2005Ap.....48..135S} {48, 135}

\bibitem[\protect\citeauthoryear{{Sitnova}, {Mashonkina}  \&
  {Ryabchikova}}{{Sitnova} et~al.}{2018}]{Sitnova2018}
{Sitnova} T.~M.,  {Mashonkina} L.~I.,   {Ryabchikova} T.~A.,  2018, \mn@doi
  [\mnras] {10.1093/mnras/sty810}, \href
  {https://ui.adsabs.harvard.edu/abs/2018MNRAS.477.3343S} {477, 3343}

\bibitem[\protect\citeauthoryear{{Siwak} et~al.,}{{Siwak} et~al.}{2018}]{Siw18}
{Siwak} M.,  et~al., 2018, \mn@doi [\mnras] {10.1093/mnras/sty1220}, \href
  {https://ui.adsabs.harvard.edu/abs/2018MNRAS.478..758S} {478, 758}

\bibitem[\protect\citeauthoryear{{Steenman} \& {The}}{{Steenman} \&
  {The}}{1989}]{ST89}
{Steenman} H.,  {The} P.~S.,  1989, \mn@doi [\apss] {10.1007/BF00650082}, \href
  {https://ui.adsabs.harvard.edu/abs/1989Ap&SS.159..189S} {159, 189}

\bibitem[\protect\citeauthoryear{{Steenman} \& {The}}{{Steenman} \&
  {The}}{1991}]{ST91}
{Steenman} H.,  {The} P.~S.,  1991, \mn@doi [\apss] {10.1007/BF00644862}, \href
  {https://ui.adsabs.harvard.edu/abs/1991Ap&SS.184....9S} {184, 9}

\bibitem[\protect\citeauthoryear{{Villebrun} et~al.,}{{Villebrun}
  et~al.}{2019}]{Vil19}
{Villebrun} F.,  et~al., 2019, \mn@doi [\aap] {10.1051/0004-6361/201833545},
  \href {https://ui.adsabs.harvard.edu/abs/2019A&A...622A..72V} {622, A72}

\bibitem[\protect\citeauthoryear{{Waters} \& {Waelkens}}{{Waters} \&
  {Waelkens}}{1998}]{WW98}
{Waters} L.~B.~F.~M.,  {Waelkens} C.,  1998, \mn@doi [\araa]
  {10.1146/annurev.astro.36.1.233}, \href
  {https://ui.adsabs.harvard.edu/abs/1998ARA&A..36..233W} {36, 233}

\makeatother
\end{thebibliography}

\bsp    
\label{lastpage}
\end{document}